# The MAterials Simulation Toolkit (MAST) for Atomistic Modeling of Defects and Diffusion

Tam Mayeshiba[a]
Henry Wu[a]
Thomas Angsten[a,1]
Amy Kaczmarowski[a,2]
Zhewen Song[a]
Glen Jenness[a,3]
Wei Xie[a,1]
Dane Morgan[a]*, ddmorgan@wisc.edu

[a]Department of Materials Science and Engineering, University of Wisconsin-Madison, 1509 University Ave., Madison, Wisconsin, 53706

[1]Present address: Department of Materials Science and Engineering, University of California-Berkeley, 210 Hearst Mining Building, Berkeley, California, 94720-1760
[2]Present address: Sandia National Laboratories, Albuquerque, New Mexico, 87185
[3]Present address: Catalysis Center for Energy Innovation, University of Delaware, 221 Academy Street, Newark, Delaware, 19716

## Abstract

The MAterials Simulation Toolkit (MAST) is a workflow manager and post-processing tool for *ab initio* defect and diffusion workflows. MAST codifies research knowledge and best-practices for such workflows, and allows for the generation and management of easily modified and reproducible workflows, where data is stored along with workflow information for data provenance tracking. MAST is available for download through the Python Package Index, or at https://pypi.python.org/pypi/MAST, with installation instructions and a detailed user's guide at http://pythonhosted.org/MAST. MAST code may be browsed at the GitHub repository at https://github.com/uw-cmg/MAST.

## 1. Introduction

Large scale automated ab initio (or first-principles) atomic scale computer simulation has a growing presence in materials science, with scientists exploring calculated materials properties over different compositions and crystal structures in series of high-throughput calculations.[1-6]

Calculating materials properties often requires a workflow of several calculations. For example, a typical workflow to arrive at an *ab initio* diffusion coefficient requires over a dozen separate density-functional theory (DFT) calculations, plus a handful of intermediate steps and additional post-processing. Automating such workflows reduces the amount of user attention necessary during the workflow and increases the uniformity among similar workflows.

Several workflow managers exist, including the Materials Project's MPWorks or Fireworks code,[1,7] the AFLOW library,[8] AiiDA,[9] the UW-Madison Center for High-Throughput Computing (CHTC) DAGMan,[10] and the qmpy software behind the Open Quantum Materials Database.[11] However, the codification of materials science knowledge and logic within a workflow is often highly tuned to a specific materials science problem. This knowledge includes calculation parameters, structural information, and links between different stages and types of calculations. Therefore, there is still a significant need to build new or extend existing tools to efficiently tackle specific problems.

The MAterials Simulation Toolkit (MAST) is an automated high-throughput workflow manager and suite of post-processing tools designed to facilitate atomic simulation calculations for defect and diffusion calculations, especially using density functional theory (DFT) as implemented by the Vienna Ab-initio Simulation Package (VASP).[12]

The goals of MAST are to codify knowledge in workflows, provide a general, flexible framework that allows for easy modification, and manage thousands of DFT workflows for research needs, without any emphasis on total automation between different types of workflows, e.g. between a k-point mesh convergence workflow and a diffusion workflow. Every existing feature and enhancement of the present tool was prompted by a research need. Where the complexity of programming some automation was not supported by a commensurate research need, such programming was not undertaken. MAST is under continuous evolution as research needs change and from usage feedback. Information on accessing MAST is given at the end of Sec. 5.

## 2. Overview

### 2.1. Operation

Figure 1 shows MAST's two primary operations: 1) creating a workflow using an input file and 2) managing workflows.

In order to create the directories and metadata for a workflow, a user would first either modify an existing input file or write one from scratch following the documentation. Then, the user would run the command **mast –i <input file>**. This command is completely separate from any management operations, so the user may input additional workflows at the same time as MAST is managing existing workflows.

In order to manage workflows, a user would run the command **mast** as often as necessary, typically once every few hours or once every day for workflows containing DFT calculations of the size and type for which MAST is intended. The user could also set up a *crontab* for the cluster to run the command periodically to allow the runs to proceed automatically.

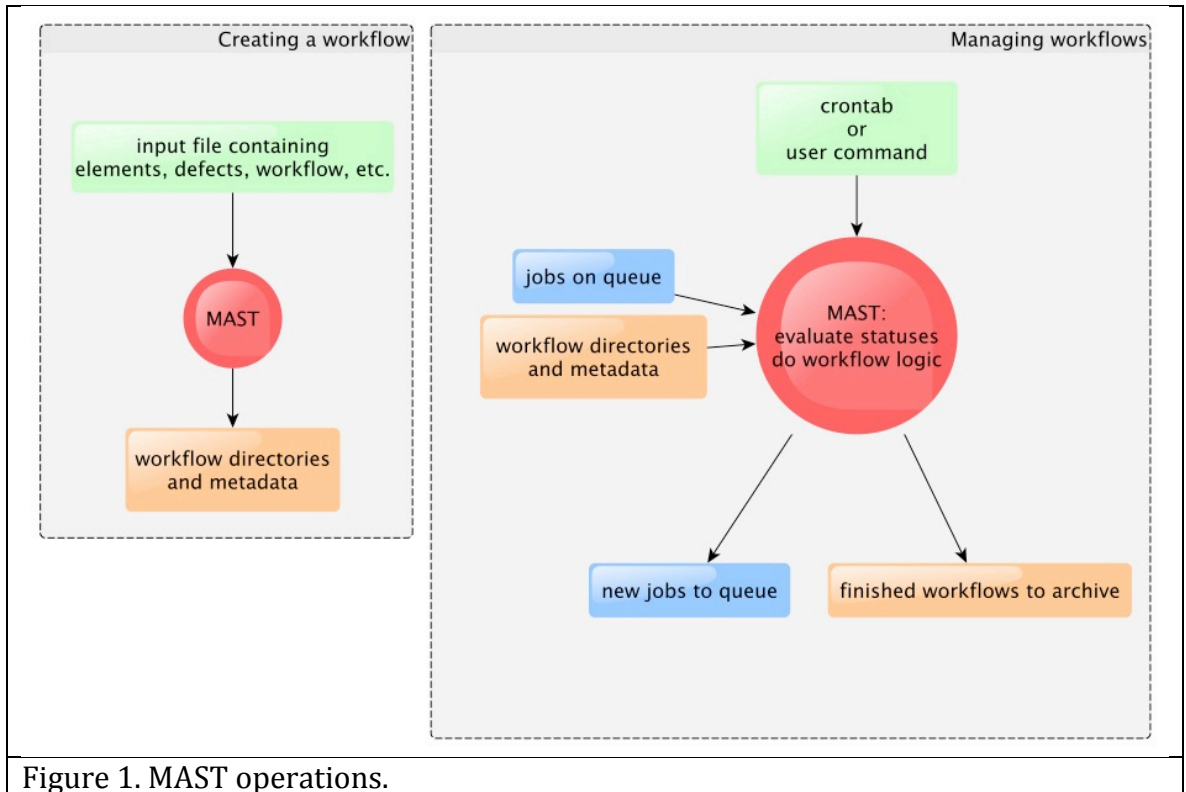

Figure 1. MAST operations.

### 2.1.1. Creating a workflow

The MAST input file is designed to allow workflows to be reproducible and easily modified. Example out-of-the-box workflows which are available in the MAST package include (1) a full impurity in a host diffusion coefficient workflow template, with consideration of phonons, multiple hops, and support for various multi-frequency models for diffusion and (2) a full defect formation energy workflow template, including support for creating charged defects, performing finite-size scaling corrections, and evaluating potential alignment for charged supercells. See Section 3 for more information on these workflows. All calculation parameters and job-to-job relationships are stored in the input file, saving the research workflow so that it can be followed later or reused with only minor modifications to previously-recorded parameters.

The MAST input file allows for extensive workflow customization. Users can easily generate complex workflows, where many successive and interconnected steps may be combined through a directed acyclic graph[13] (DAG) format. That is, jobs may wait for the completion of any number of other "parent" jobs, and in turn may pass data to any number of other "child" jobs.

MAST also allows additional workflow branches to be appended to existing workflows both mid-completion and after completion, for example, if the user decides that extra charge states, additional supercell sizes, or additional defects and hops are desirable. These custom workflows are all saved for future use, making it easy to reproduce results and to apply identical workflows to different systems.

A looping feature in the input file allows the same input file to be easily used for different elements or parameter combinations. Other notable features of MAST include the ability to apply lattice strain, a Madelung energy-based utility for picking supercell sizes,[14] and atomic indexing to keep track of individual atoms.

See Appendix B for a rudimentary input file example.

### 2.1.2. Managing workflows

Figure 1 shows that MAST collects queue status information and information from the workflow directories, including output text files and metadata text files, in order to manage workflows. MAST uses this information to check the progress of individual jobs in the workflow, set up the next jobs when parent jobs complete, and check the overall completion status of the workflow.

Through a combination of MAST's internal error handlers and pymatgen's custodian error handlers,[15] MAST can detect, fix, and resubmit jobs that encounter some common errors, while allowing all other independent calculation branches to proceed normally. If MAST corrects an error, the pre-error calculation files are saved in a zipped archive file in the calculation directory, and an error trail is preserved in the calculation directory.

MAST manages workflows independently from each other, so that any particular workflow may be skipped or fail due to errors without affecting other workflows. MAST can also be interrupted and rerun without consequences, because its decision-making is only informed by text files that reside on the cluster, and not from any persistent stored memory of its own.

Completed workflows are automatically moved to a separate folder so that users can easily tell which workflows have completed.

### 2.1.3. Handling completed workflow data

MAST employs workflow identification features that make it possible for users at a later date to make sense of archived research data. The foremost identification feature is the input file, which shows parent/child job relationships through an indented DAG structure, and input parameter information for each calculation type. When MAST is allowed to tag-generate calculation names, these names themselves are designed to make identification easier, including for example the defect name, the scaling size, the charge on the supercell, and/or the hop name. The standardization of calculation names also makes it easy to write additional post-processing scripts for data.

MAST also generates a Bibtex-format citation file in the workflow directory of completed workflows. This file includes citations for all major methods and externally contributed pieces of code, and allows researchers to easily import citations into publication drafts and to identify which methods were used in the workflow.

### 2.2. Portability

MAST has been designed to work over external shared-resource computing clusters in order to maximize computing opportunities. Therefore, MAST has the following features:
- MAST is compatible with several major resource management systems, including SLURM, SGE, PBS/Torque, and even HTCondor with no shared home directory, and can be adapted to more.
- All intensive MAST monitor processes (see the red circle under Managing Workflows in Figure 1) run as a serial job over the queue system, so that MAST has low headnode load appropriate for shared resources. The MAST monitor is neither a continuous nor a headnode process.
- The installation of MAST does not require root access, so any user can install and use MAST on an external cluster with a user account. MAST also uses the Python language (as opposed to C++ or Fortran), and it is relatively easy to set up a user-based Python 2.7.X installation with the required python package dependencies.

### 2.3. Stability and Extensibility

Many key MAST functions depend on the Materials Project's pymatgen library (http://pymatgen.org), drawing stability from pymatgen's testing resources and user base and avoiding duplication of effort.

Also, MAST programmers can add citations for functions and pieces of code to the MAST automatic citation mechanism in order to encourage collaboration and the inclusion of additional useful functions.

Through a modular program-compatibility structure, common algorithms like transferring a relaxed structure to a new calculation may be redefined for different programs without changing any higher program architecture. For example, the currently supported program is VASP, but an identical module could be set up for a program like ABINIT.[16] Also, the modular structure optimization library construction in the StructOpt package allows advanced algorithms to be programmed into MAST in order to extend beyond its present genetic algorithm approaches.

## 2.4. Summary and future outlook

MAST is:
- A low-overhead, no-root-access-required workflow manager that works with many different cluster setups and produces organized output.
- A source for out-of-the-box workflows for defect and diffusion calculations, including the logic to specify which information goes to which other calculations, and which parameter setup to use for each calculation.
- Programmed to support common workflow elements for defect and diffusion calculations, such as the ability to scale supercells and defects, loop over parameters and elements, apply charge to defects, and set up and run nudged elastic band calculations.

MAST is not:
- An all-in-one data design environment: if not using one of MAST's out-of-the-box workflows, a user must already have an idea of the scope and parameters of the calculations to be run, and to be able to map out the workflow in order to transform it into an input file.
- An automatic database generator: MAST produces organized input and output, making it easy to search for, find, track, and identify data, but at present it does not automatically create a formal database out of its input or output.
- Capable of workflows beyond DAGs: MAST can perform any predefined DAG structured workflow. However, MAST cannot perform workflows with open-ended loops or decisions. More specifically, MAST cannot decide on a workflow branch based on the results of a previous calculation, or produce an open-ended number of calculations based on a criterion.

Future MAST development is expected to concentrate on providing support for additional defect- and diffusion-related functions and workflows, as well as to provide support for Python3. Workflow management development is expected to be limited to minor usability issues and increased compatibility with existing workflow managers, for example, possibly supporting the Fireworks[7] DAG in addition to the supported HTCondor[10] DAG. Automated database development is expected to be minimal and to follow the Materials Project's MPContribs framework.[17]

## 3. Sample workflows

This section details two major sample workflows packaged with MAST: the diffusion coefficient workflow (Section 3.1), and the defect formation energy workflow, (Section 3.2). In practice, MAST workflows usually encompass some 10-20 separate *ab initio* calculations, with each workflow being applied to some tens to hundreds of systems.

This section assumes basic familiarity with *ab initio* density functional theory calculations using VASP, and its associated terminology.

### 3.1. Diffusion coefficient

The MAST diffusion coefficient workflow sets up and runs VASP climbing image nudged elastic band (CI-NEB) calculations[18,19] in order to calculate an impurity diffusion coefficient over a range of temperatures. Figure 2 shows the diffusion coefficient workflow for both the pure solvent and an impurity solute on a face-centered cubic lattice using the five-frequency model of diffusion[20,21] (The complete workflow integrated into one figure can be found in Figure A.1.)

In this example workflow, the desired result is the vacancy-mediated diffusivity of a solute element within a host element. In Figure 2b, working from the bottom up, we see that diffusivity is obtained from the dilute diffusion post-processing utility (gray box), which takes in several different frequencies (purple box). These frequencies each correspond to rates of different atomic hops in the vicinity of the solute atom and host vacancy. An atomic hop is characterized by its activation barrier, calculated by NEB (green box), and its attempt frequency, calculated using phonons (blue box). Both NEB and phonon calculations depend upon relaxed defect structures (yellow box) to serve as initial configurations. Finally, all defect structures are defined by and generated from (orange box) the perfect and unperturbed host element lattice (red box).

Each defect calculation and nudged elastic band calculation may be run at any charge state. Variations on this workflow have been used in several publications.[5,6,22]

a) 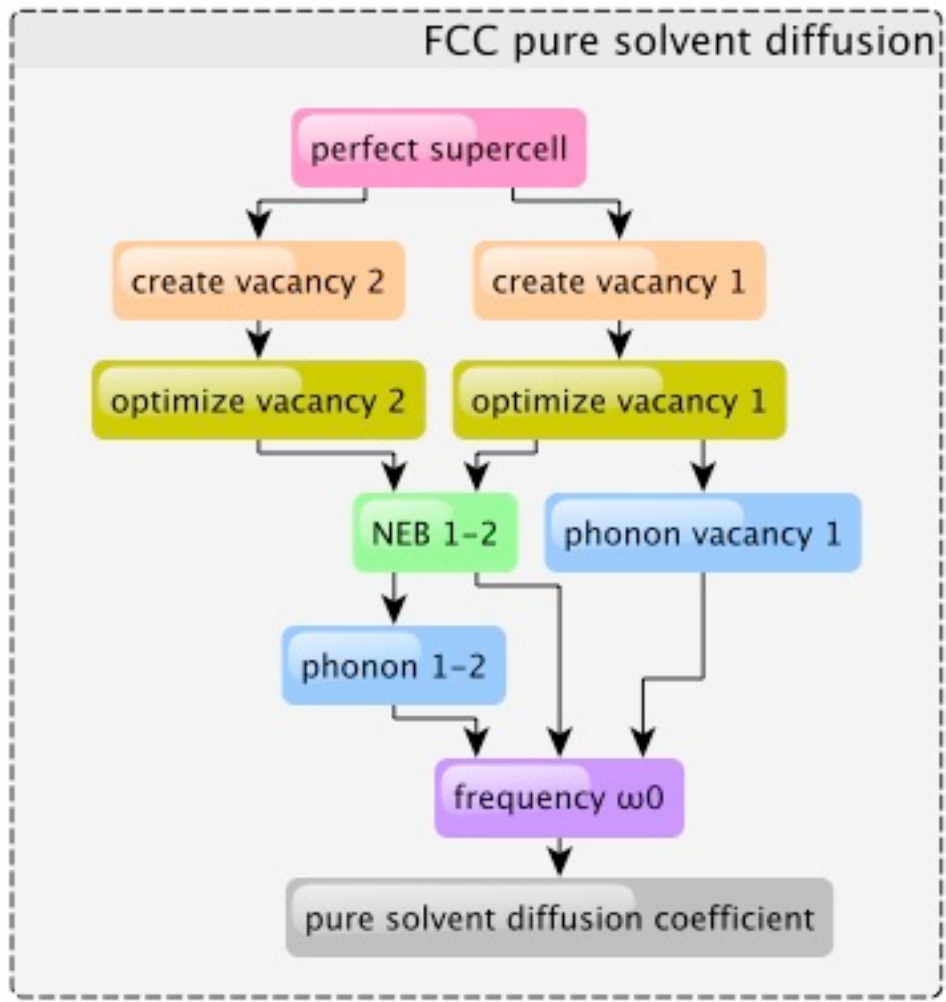

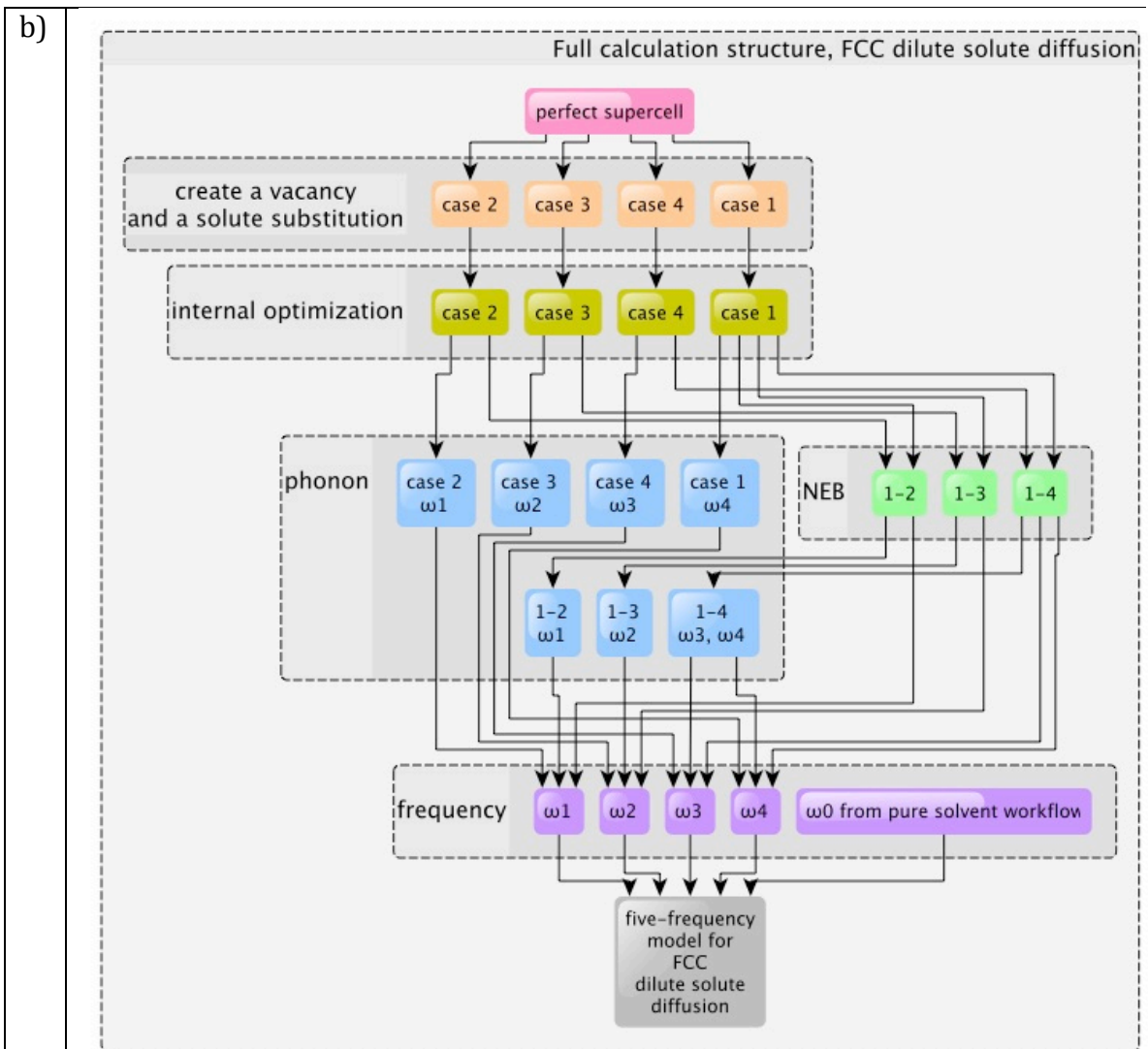

Figure 2. Diffusion coefficient workflow for (a) the pure solvent and (b) an impurity solute. See Figure A.1 for the full, integrated impurity solute diffusion coefficient workflow.

Figure 3 shows diffusion coefficient versus temperature data from several diffusion coefficient workflows plotted on the same axes. Because the data for each host was calculated consistently using the same workflow, the data is readily comparable.

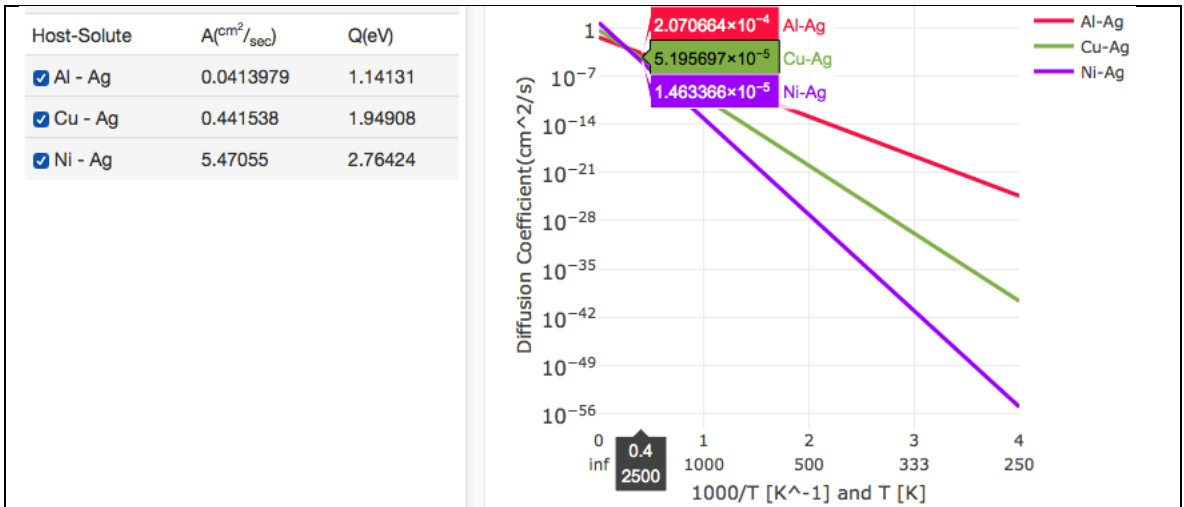

Figure 3. Sample output from a diffusion coefficient workflow, plotted. Screenshot is from the UW CMG Dilute Solute Diffusion Database.[23]

Figure 4 shows diffusion barriers plotted versus the solute atom in host FCC Aluminum. With the automated capabilities of MAST, it becomes a simple matter of stepping through the periodic table to perform these studies, and only by consistent calculations can trends be so easily discerned. We can see from Figure 4 that, excepting the magnetic elements, diffusion of transition metal solutes in Al increases with the addition of d-electrons until it reaches a peak near the middle of the d-block. Then the diffusion barrier decreases as we continue to the right of the periodic table and plateaus for the post-transition elements.

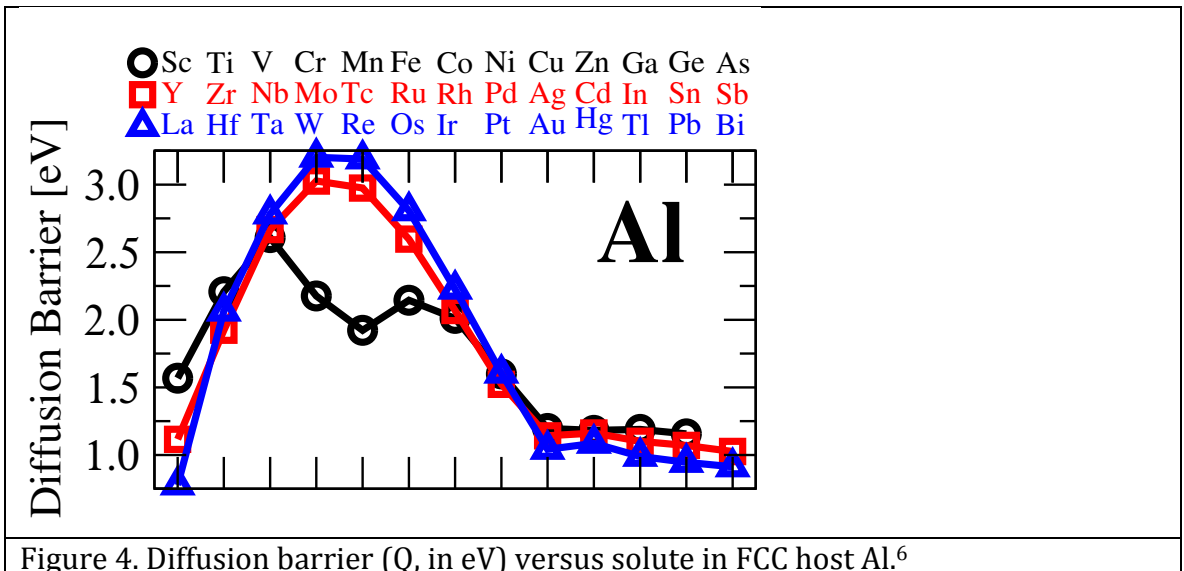

Figure 4. Diffusion barrier (Q, in eV) versus solute in FCC host Al.[6]

## 3.2. Defect formation energy

The defect formation energy workflow calculates isolated (or highly-dilute non-interacting) defect formation energies over user-input chemical potential conditions and can be found in Figure 5. (The complete workflow integrated into one figure can be found in Figure A.2.) It handles finite-size scaling corrections,[24-26] potential alignment corrections,[27] and an approach to incorporate hybrid DFT energy corrections.[28]

In the example workflow from Figure 5, we show our procedure to determine formation energies for charged defects corrected using the Heyd-Scuseria-Ernzerhof (HSE) functional.[29-31] To begin, a list of the defects of interest and their relevant range of charges are needed. Because non-neutral defects exhibit long-range electrostatic interactions (and potentially long-range strain interactions), the errors introduced by finite-size simulation cells need to be corrected. We correct for these finite-size errors by computing all defects and all charges with several supercell sizes (purple boxes), which are taken together in the post-processing utility to extrapolate the defect energy at infinite-size following the finite-size scaling approach of Castleton et al.[32]

A separate difficulty in comparing differently-charged defects is the problem of slightly different electron reference energies. We correct for this effect by computing the potential alignment shift between all defects and charges with respect to the undefected, perfect supercell. While there are many methods to compute the potential alignment shift, we utilize the mean shift from all host atoms, which has been shown to converge quickly and in a linear fashion.[27,33]

Lastly and perhaps most importantly, the accuracy of theoretical defect formation energies calculated from many common exchange-correlation functionals is highly uncertain. This is an intrinsic limitation, showing up most notably in the significant underestimation of band gaps from both local density approximation (LDA) and generalized gradient approximation (GGA). We address this limitation by applying a correction derived from a hybrid Hartree-Fock functional (HSE[29]). While a user could simply run the full calculation with HSE or a comparable functional, these more accurate functions typically greatly increase the computational time to the point that running many well-converged and/or large cells for finite-size scaling analysis becomes impractical. To circumvent this problem we use a mixed GGA and HSE scheme developed by Jacobs, et al.[28] and validated on the $Sc_2O_3$ system. In this scheme, an energy difference is computed for all defects and charges between GGA and HSE, at both a small supercell size and a decreased k-point mesh (yellow boxes). We then assume this energy difference is a constant shift between GGA and HSE for the particular defect/charge for any cell size and k-point mesh and apply the same shift for all of our GGA calculations, adding them on top of preceding finite-size and potential alignment corrections.

a)
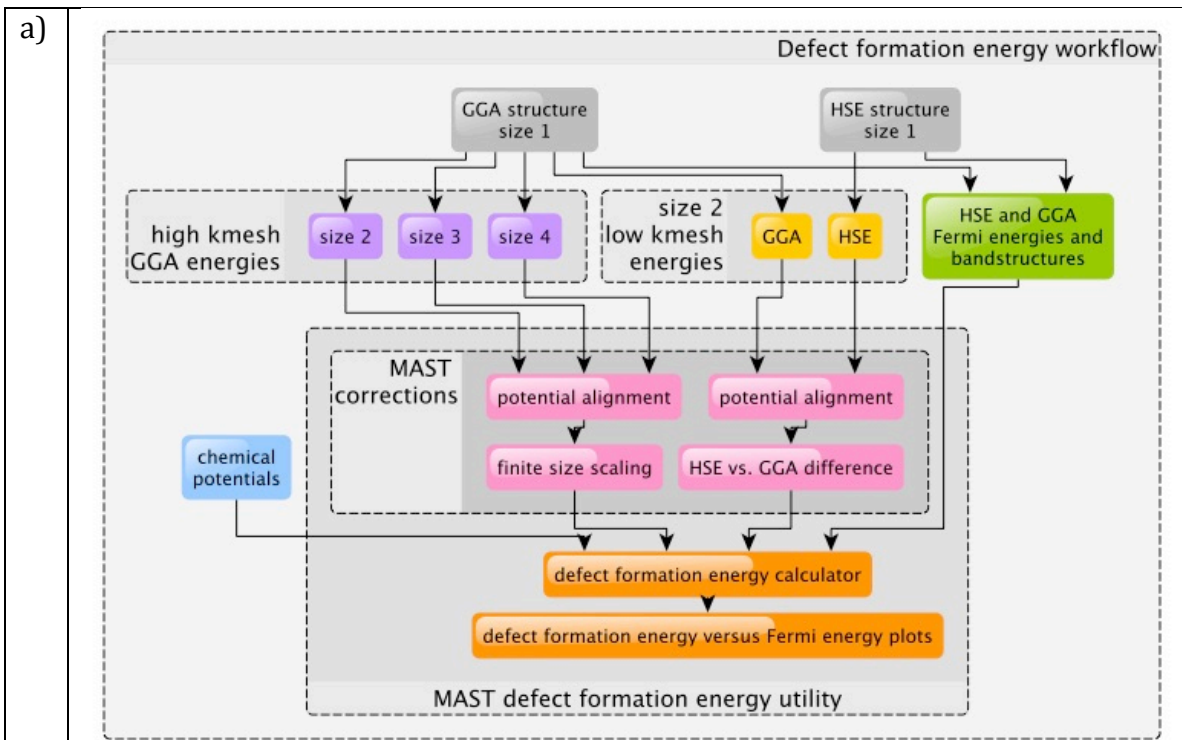

b)
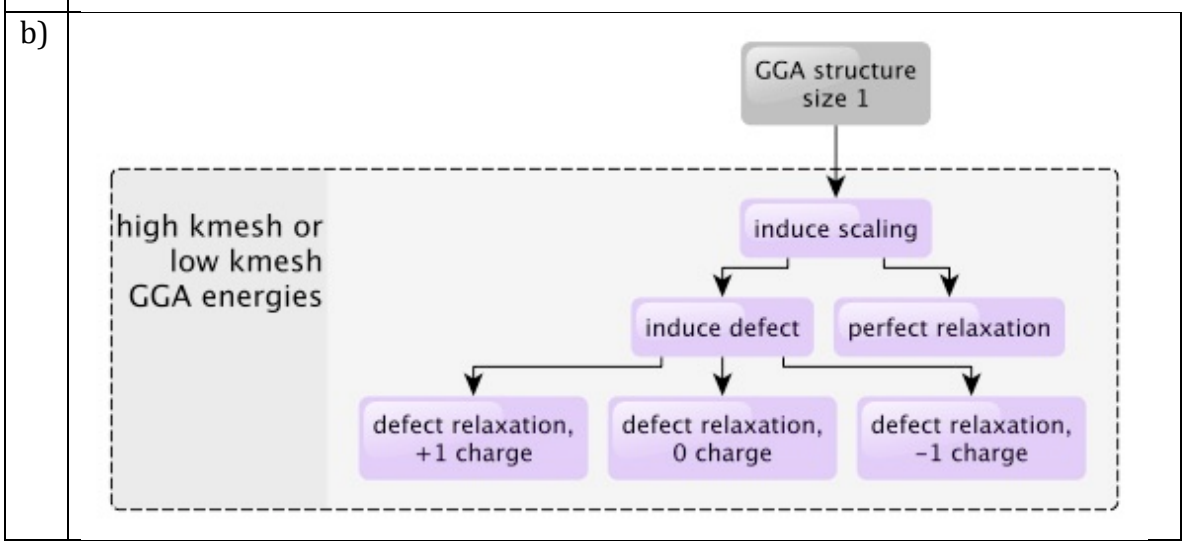

| | |
|---|---|
| c) | 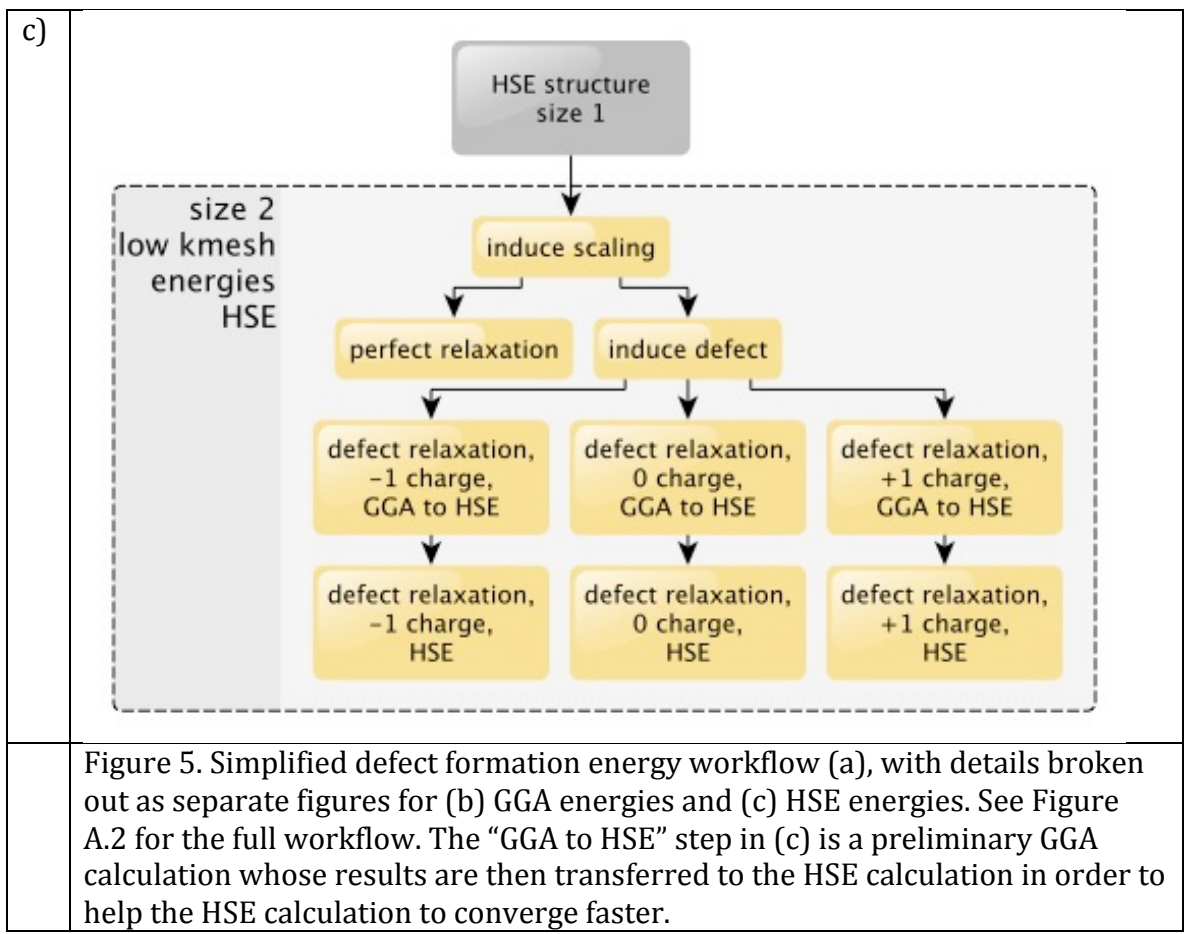 |
| | Figure 5. Simplified defect formation energy workflow (a), with details broken out as separate figures for (b) GGA energies and (c) HSE energies. See Figure A.2 for the full workflow. The "GGA to HSE" step in (c) is a preliminary GGA calculation whose results are then transferred to the HSE calculation in order to help the HSE calculation to converge faster. |

Figure 6 shows sample output from the defect formation energy workflow for GaN and GaP. All desired defects and charge states are automatically evaluated together using the workflow from Figure 5. All finite-size, potential alignment, and HSE shift corrections have been automatically applied.

Symmetry-distinct defect sites may be found using the MAST Defect Finder, which identifies occupied sites for vacancies, grids 3-D space to create dumbbells, and creates interstitials through gridding of Wyckoff positions and at the centers, edge centers, and face-centers of Voronoi-tessellated tetrahedra. The defect finder is available upon request and will be packaged with later versions of MAST.

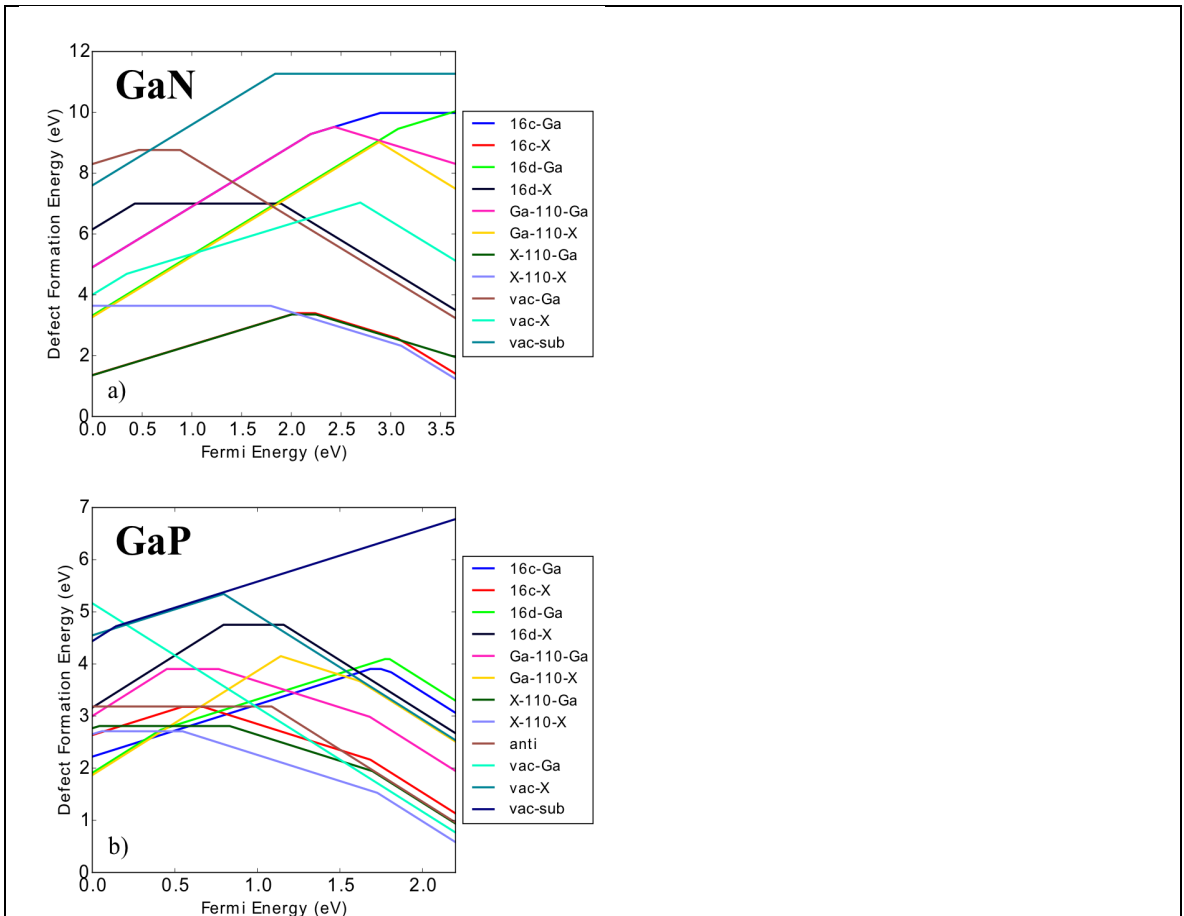

Figure 6. HSE defect formation energies for intrinsic defects in a) GaN and b) GaP. In each legend, "X" stands for the elements N or P. "16c-Ga" stands for a Ga interstitial at Wyckoff position 16c, and so on. "Ga-110-Ga" stands for a Ga-Ga dumbbell in the <110> direction, and so on. "vac-Ga" stands for a Ga vacancy, and so on. "anti" is a $Ga_P$ antisite with a nearest-neighbor (NN) $P_{Ga}$ antisite. "vac-sub" is a Ga vacancy with a NN $Ga_X$ antisite. In these plots, the chemical potentials $\mu$ are taken under Ga-rich conditions, such that $\mu(Ga) = E(Ga)$ calculated from crystal Ga DFT energies and $\mu(X) = E(GaX) – E(Ga)$.

## 4. Additional features

MAST also includes several stand-alone packages. The foremost of these packages is the StructOpt genetic algorithm structure optimization code,[34] which operates in both a stand-alone and an integrated workflow-managed manner. Instead of having the user create a few dozen arrangements by hand, StructOpt can evaluate thousands of arrangements with a genetic algorithm search. StructOpt can optimize structures to find those with lowest energy for crystals, molecules, nanoparticles, and defects within a crystal host. StrucOpt also runs with the Large-scale

Atomic/Molecular Massively Parallel Simulator (LAMMPS)[35] software for optimization with interatomic potential Hamiltonians.

## 5. Conclusions

MAST has proven to be a useful workflow management and post-processing tool for defect and diffusion calculation, facilitating tens of thousands of research calculations and millions of CPU-hours worth of defect and diffusion data. Data produced by MAST was highly reproducible and identifiable, with clear provenance tracking. The standardized calculation naming resulted in data that was easy to search, and consistently formatted for post-processing. Input files were easy to modify and reuse from one project to another, or within the same project in order to perform additional calculations.

In conclusion, MAST allows for the setup and management of consistent, reproducible *ab-initio* diffusion and defect workflows. By enabling automated high-throughput *ab initio* calculations, MAST facilitates the creation of databases for mining and further acceleration of materials development. Publications using MAST to date include Refs [5,6,22,34,36-39].

For anyone interested in using MAST, a complete user's guide is available at http://pythonhosted.org/MAST, with the code in the Python Package Index at http://pypi.python.org/pypi/MAST and also on github at http://github.com/uw-cmg/MAST. Older code versions are archived on the Zenodo repository at http://zenodo.org. MAST is released under the terms of the MIT license.[40]

## Acknowledgments

MAST development takes place in the Computational Materials Group (CMG) at the University of Wisconsin-Madison as part of the NSF Software Infrastructure for Sustained Innovation (SI[2]) program funded by NSF award number 1148011, and has enjoyed technical collaboration with developers of pymatgen[15] and the Materials Project (MP).[1]

T. Mayeshiba gratefully acknowledges support from the NSF Graduate Fellowship Program under Grant No. DGE-0718123 and an Advanced Opportunity Fellowship through the Graduate Engineering Research Scholars program at the University of Wisconsin-Madison. Many underlying MAST functions are built using pymatgen (http://pymatgen.org), and the MAST team would especially like to thank pymatgen developers Dr. Shyue Ping Ong and Dr. Anubhav Jain for their assistance.

Many people have been involved in the creation and testing of MAST, and they are listed below in alphabetical order according to last name. Major contributors are marked with an asterisk (*), those who have undertaken research performed using

MAST are marked with a plus sign (+), and where an individual's contribution is limited to a specific contribution, it is noted in parentheses.

PI: Professor Dane Morgan

Benjamin Afflerbach (fall 2014 - present)
Nada Alameddine (summer 2013, Defect Finder)
Thomas Angsten +* (spring 2011 - summer 2013)
Dr. Leland Barnard (spring 2014, Particle Diffusion Trajectory Analysis)
Jesus Chavez (summer 2014)
Saswati De (summer 2014)
Dr. Jie Deng (spring 2014, Effective Grain Boundary Diffusivity Calculator)
Dr. Ryan Jacobs + (testing and suggestions)
Dr. Glen Jenness +* (spring 2013 - summer 2013)
Chandana Hosamane Kabbali (summer 2014)
Dr. Anubhav Jain (suggestions)
Amy Kaczmarowski +* (fall 2013 – spring 2014; StructOpt)
Hyunwoo Kim (spring 2013)
Daniil Kitchaev (NEB pathfinder[41])
Hyunseok Ko + (StructOpt testing)
Dr. Guangfu Luo (suggestions)
Tam Mayeshiba +* (summer 2010 - present)
Kumaresh Visakan Murugan * (spring 2013, fall 2013, spring 2014)
Jihad Naja (summer 2013, Defect Finder)
Dr. Hyo On Nam + (testing and suggestions)
Parker Sear (spring 2013 - summer 2013, spring 2014)
Zhewen Song +* (fall 2013 – spring 2016)
Dr. Henry Wu +* (summer 2013 - present)
Dr. Wei Xie (fall 2013 – fall 2014)
Dr. Min Yu +* (StructOpt testing)

# Appendices

## Appendix A. Workflow diagrams

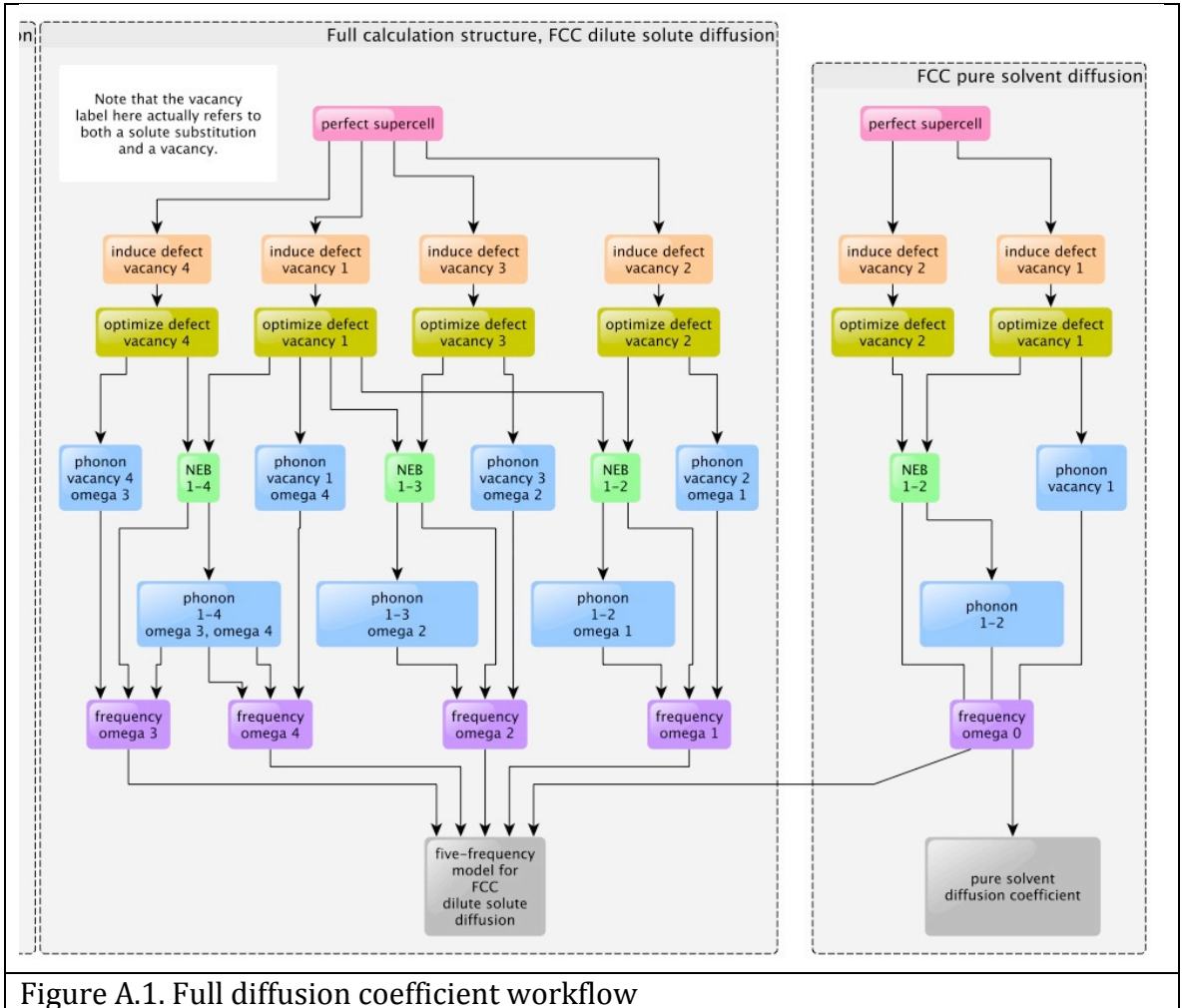

Figure A.1. Full diffusion coefficient workflow

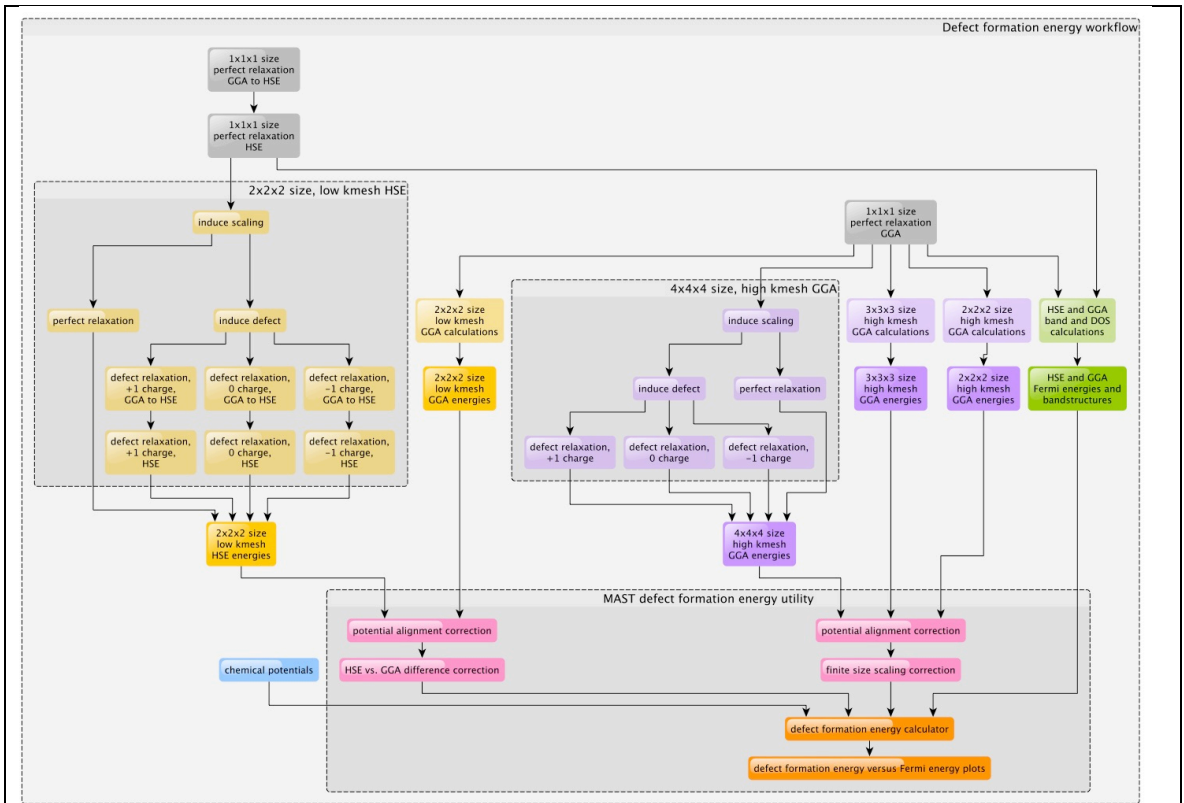
Figure A.2. Full defect formation energy workflow

For Figure A.2, the chemical potentials are entered as user input in the input file. Because there are many choices involved in obtaining the chemical potential (e.g. choice of reference state/phase, calculation type, and using experimental values instead of calculated values), MAST does not automate a calculated chemical potential into the defect workflow.

Workflow diagrams were generated using the yEd Graph Editor.[42]

## Appendix B. Input file example

Table B.1 annotates a workflow that calculates multiple defect energies (Al vacancy, Mg interstitial, and Cu substitution) in an FCC Al system. The input file is written in plain text with $<section name> and $end to differentiate sections. It was designed to be easy for humans to produce and read. This input file is a simple example. For additional input file sections and explanation, especially for features associated with the workflows in Appendix A, please see the user documentation at http://pythonhosted.org/MAST.

Table B.1. Simple optimization workflow input file, annotated.

| `$mast` | |
|---|---|
| | "$<section name>" to begin a section |
| `system_name OptimizeWorkflowTest` | System name gets prepended to the workflow folder |
| `$end` | "$end" to end a section |
| | |
| `$structure` | |
| `use_structure_index True` | Keep track of atoms no matter where they move (beta) |
| `coord_type fractional` | |
| | |
| `begin elementmap` | |
| `X1 Al` | Using X1 in the rest of the input file allows the user to copy this input file, switch the element labels here, and reuse most of the rest of the input file without modification. |
| `X2 Mg` | |
| `X3 Cu` | |
| `end` | |
| | |
| `begin lattice` | |
| `7.0 0 0` | |
| `0 7.0 0` | |
| `0 0 7.0` | |
| `end` | |
| | |
| `begin coordinates` | |
| `X1 0.0000000000 0.0000000000 0.0000000000` | |
| `X1 0.0000000000 0.0000000000 0.5000000000` | |
| `X1 0.0000000000 0.5000000000 0.0000000000` | |
| `X1 0.0000000000 0.5000000000 0.5000000000` | |
| `X1 0.5000000000 0.0000000000 0.0000000000` | |
| `X1 0.5000000000 0.0000000000 0.5000000000` | |
| `X1 0.5000000000 0.5000000000 0.0000000000` | |
| `X1 0.5000000000 0.5000000000 0.5000000000` | |
| `X1 0.2500000000 0.2500000000 0.0000000000` | |
| `X1 0.2500000000 0.2500000000 0.5000000000` | |
| `X1 0.2500000000 0.7500000000 0.0000000000` | |
| `X1 0.2500000000 0.7500000000 0.5000000000` | |
| `X1 0.7500000000 0.2500000000 0.0000000000` | |
| `X1 0.7500000000 0.2500000000 0.5000000000` | |
| `X1 0.7500000000 0.7500000000 0.0000000000` | |
| `X1 0.7500000000 0.7500000000 0.5000000000` | |
| `X1 0.0000000000 0.2500000000 0.2500000000` | |
| `X1 0.0000000000 0.2500000000 0.7500000000` | |
| `X1 0.0000000000 0.7500000000 0.2500000000` | |
| `X1 0.0000000000 0.7500000000 0.7500000000` | |
| `X1 0.5000000000 0.2500000000 0.2500000000` | |
| `X1 0.5000000000 0.2500000000 0.7500000000` | |
| `X1 0.5000000000 0.7500000000 0.2500000000` | |
| `X1 0.5000000000 0.7500000000 0.7500000000` | |
| `X1 0.2500000000 0.0000000000 0.2500000000` | |
| `X1 0.2500000000 0.0000000000 0.7500000000` | |
| `X1 0.2500000000 0.5000000000 0.2500000000` | |
| `X1 0.2500000000 0.5000000000 0.7500000000` | |
| `X1 0.7500000000 0.0000000000 0.2500000000` | |
| `X1 0.7500000000 0.0000000000 0.7500000000` | |
| `X1 0.7500000000 0.5000000000 0.2500000000` | |
| `X1 0.7500000000 0.5000000000 0.7500000000` | |
| `end` | |
| | |
| `$end` | |
| | |
| `$defects` | |
| `begin vac1` | "vac1" is the name for this defect grouping |
| `vacancy 0.0 0.0 0.0 X1` | This defect grouping contains a single vacancy defect of type X1, which corresponds to Al. |
| `end` | |
| | |
| `begin int1` | |
| `interstitial 0.125 0.125 0.125 X2` | |
| `end` | |
| | |
| `begin sub1` | |
| `substitution 0.25 0.25 0.0 X3` | |
| `end` | |
| | |
| `$end` | |
| | |
| `$ingredients` | |
| `begin ingredients_global` | This section specifies defaults for the |

|  |  | calculations, whose overrides will be in subsequent sections. |
|---|---|---|
| `###Change the options below for your cluster` |  |  |
| `mast_nodes          1` |  | mast_XXX keywords are special program keywords |
| `mast_multiplyencut 1.5` |  |  |
| `mast_ppn            1` |  |  |
| `mast_queue          default` |  |  |
| `mast_exec           //share/apps/vasp5.2_cNEB` |  |  |
| `mast_walltime       4` |  |  |
| `###` |  |  |
| `mast_kpoints        2x2x2 M` |  |  |
| `mast_xc             PBE` |  |  |
| `isif 3` |  | Program-specific keywords are listed here. The keywords in this example are for VASP, and will produce a calculation's INCAR file. |
| `ibrion 2` |  |  |
| `nsw 191` |  |  |
| `ismear 1` |  |  |
| `sigma 0.2` |  |  |
| `lwave False` |  |  |
| `lcharg False` |  |  |
| `prec Accurate` |  |  |
| `mast_program    vasp` |  |  |
| `mast_write_method` | `write_singlerun` | MAST control keywords |
| `mast_ready_method` | `ready_singlerun` |  |
| `mast_run_method` | `run_singlerun` |  |
| `mast_complete_method` | `complete_singlerun` |  |
| `mast_update_children_method` | `give_structure` |  |
| `end` |  |  |
|  |  |  |
| `begin inducedefect` |  | This type of ingredient has overrides for certain global ingredient keywords. |
| `mast_write_method` | `no_setup` |  |
| `mast_ready_method` | `ready_structure` |  |
| `mast_run_method` | `run_defect` |  |
| `mast_complete_method` | `complete_structure` |  |

| | |
|---|---|
| **end** | |
| | |
| **begin lowmesh** | |
| **mast_kpoints 1x1x1 G** | |
| **end** | |
| | |
| **begin static** | |
| **ibrion -1** | |
| **nsw 0** | |
| **mast_multiplyencut 1.25** | |
| **end** | |
| **$end** | |
| | |
| **$recipe** | The Recipe section is the workflow section. Child ingredients are indented. |
| **perfect_opt1 (lowmesh)** | The ingredient's type, from the ingredient section, is denoted in parentheses. |
|    **perfect_opt2** | No parentheses means the ingredient just gets global values. |
|      **perfect_stat (static)** | |
|      **inducedefect_<N> (inducedefect)** | <N> tags will draw from the defect section's defect grouping names, like vac1, sub1, and int1. |
|         **defect_<N>_opt1 (lowmesh)** | |
|           **defect_<N>_opt2** | |
|             **defect_<N>_stat (static)** | |
| **$end** | |
| | |
| **$summary** | The optional summary section tells MAST to collect certain values from certain calculations after the workflow is complete. |
| **perfect energy** | |
| **vac1 energy** | |
| **int energy** | |
| **sub energy** | |
| **$end** | |

# References

1   Jain, A. *et al.* Commentary: The Materials Project: A materials genome approach to accelerating materials innovation. *APL Materials* **1**, 011002, doi:10.1063/1.4812323 (2013).
2   Curtarolo, S., Morgan, D. & Ceder, G. Accuracy of ab initio methods in predicting the crystal structures of metals: A review of 80 binary alloys. *Calphad* **29**, 163-211, doi:10.1016/j.calphad.2005.01.002 (2005).
3   Saal, J. E., Kirklin, S., Aykol, M., Meredig, B. & Wolverton, C. Materials Design and Discovery with High-Throughput Density Functional Theory: The Open Quantum Materials Database (OQMD). *Jom* **65**, 1501-1509, doi:10.1007/s11837-013-0755-4 (2013).
4   Pyzer-Knapp, E. O., Simm, G. N. & Aspuru Guzik, A. A Bayesian approach to calibrating high-throughput virtual screening results and application to organic photovoltaic materials. *Mater. Horiz.* **3**, 226-233, doi:10.1039/c5mh00282f (2016).
5   Angsten, T., Mayeshiba, T., Wu, H. & Morgan, D. Elemental vacancy diffusion database from high-throughput first-principles calculations for fcc and hcp structures. *New Journal of Physics* **16**, 015018, doi:10.1088/1367-2630/16/1/015018 (2014).
6   Wu, H., Mayeshiba, T. & Morgan, D. High-throughput ab-initio dilute solute diffusion database. *Sci Data* **3**, 160054, doi:10.1038/sdata.2016.54 (2016).
7   Jain, A. *et al.* FireWorks: a dynamic workflow system designed for high-throughput applications. *Concurrency and Computation: Practice and Experience* **27**, 5037-5059, doi:10.1002/cpe.3505 (2015).
8   Curtarolo, S. *et al.* AFLOWLIB.ORG: A distributed materials properties repository from high-throughput ab initio calculations. *Computational Materials Science* **58**, 227-235, doi:10.1016/j.commatsci.2012.02.002 (2012).
9   Pizzi, G., Cepellotti, A., Sabatini, R., Marzari, N. & Kozinsky, B. AiiDA: automated interactive infrastructure and database for computational science. *Computational Materials Science* **111**, 218-230, doi:10.1016/j.commatsci.2015.09.013 (2016).
10  *HTCondor High Throughput Computing*, <https://research.cs.wisc.edu/htcondor/manual/> (2016).
11  Wolverton, C. *OQMD: An Open Quantum Materials Database*, <oqmd.org> (2016).
12  Hafner, J., Kresse, G., Vogtenhuber, D. & Marsman, M. *Vienna Ab-initio Simulation Package*, <http://cms.mpi.univie.ac.at/vasp/, http://www.vasp.at> (
13  Martin, D. F. & Estrin, G. Experiments on Models of Computations and Systems. *IEEE Transactions on Electronic Computers* **16**, 59-69 (1967).

14    Hine, N. D. M., Frensch, K., Foulkes, W. M. C. & Finnis, M. W. Supercell size scaling of density functional theory formation energies of charged defects. *Physical Review B* **79**, 13, doi:10.1103/PhysRevB.79.024112 (2009).
15    Ong, S. P. *et al.* Python Materials Genomics (pymatgen): A robust, open-source python library for materials analysis. *Computational Materials Science* **68**, 314-319, doi:10.1016/j.commatsci.2012.10.028 (2013).
16    *ABINIT*, <abinit.org> (2016).
17    Huck, P. *et al.* User applications driven by the community contribution framework MPContribs in the Materials Project. *Concurrency and Computation: Practice and Experience* **28**, 1982-1993, doi:10.1002/cpe.3698 (2016).
18    Henkelman, G., Uberuaga, B. P. & Jónsson, H. A climbing image nudged elastic band method for finding saddle points and minimum energy paths. *The Journal of Chemical Physics* **113**, 9901, doi:10.1063/1.1329672 (2000).
19    Henkelman, G. & Jónsson, H. Improved tangent estimate in the nudged elastic band method for finding minimum energy paths and saddle points. *J Chem Phys* **113**, 9978-9985, doi:Pii [S0021-9606(00)70546-0]
Doi 10.1063/1.1323224 (2000).
20    Vineyard, G. H. Frequency factors and isotope effects in solid state rate processes. *J. Phys. Chem. Solids* **3**, 121-127 (1957).
21    Howard, R. & Manning, J. Kinetics of Solute-Enhanced Diffusion in Dilute Face-Centered-Cubic Alloys. *Physical Review* **154**, 561-568, doi:10.1103/PhysRev.154.561 (1967).
22    Mayeshiba, T. & Morgan, D. Strain effects on oxygen migration in perovskites. *Physical chemistry chemical physics : PCCP* **17**, 2715-2721, doi:10.1039/c4cp05554c (2015).
23    Boyer, E., Wu, H., Mayeshiba, T., Finkel, R. & Morgan, D. *UW CMG Dilute Solute Diffusion Database*, <http://diffusiondata.materialshub.org> (2016).
24    Lany, S. & Zunger, A. Assessment of correction methods for the band-gap problem and for finite-size effects in supercell defect calculations: Case studies for ZnO and GaAs. *Physical Review B* **78**, 17-20, doi:10.1103/PhysRevB.78.235104 (2008).
25    Makov, G. & Payne, M. Periodic boundary conditions in ab initio calculations. *Physical Review B* **51**, 4014-4022, doi:10.1103/PhysRevB.51.4014 (1995).
26    Lany, S. & Zunger, A. Accurate prediction of defect properties in density functional supercell calculations. *Model. Simul. Mater. Sci. Eng.* **17**, 14, doi:10.1088/0965-0393/17/8/084002 (2009).
27    Lin, S.-k., Yeh, C.-k., Puchala, B., Lee, Y.-L. & Morgan, D. Ab initio energetics of charge compensating point defects: A case study on MgO. *Computational Materials Science* **73**, 41-55, doi:10.1016/j.commatsci.2013.02.005 (2013).
28    Jacobs, R. M., Booske, J. H. & Morgan, D. Intrinsic defects and conduction characteristics of $Sc_{2}O_{3}$ in thermionic cathode systems. *Physical Review B* **86**, doi:10.1103/PhysRevB.86.054106 (2012).
29    Heyd, J., Scuseria, G. E. & Ernzerhof, M. Hybrid functionals based on a screened Coulomb potential. *The Journal of Chemical Physics* **118**, 8207, doi:10.1063/1.1564060 (2003).

30  Paier, J. *et al.* Screened hybrid density functionals applied to solids. *J Chem Phys* **124**, 154709, doi:10.1063/1.2187006 (2006).
31  Paier, J. *et al.* Erratum: "Screened hybrid density functionals applied to solids" [J. Chem. Phys. 124, 154709 (2006)]. *The Journal of Chemical Physics* **125**, 249901, doi:10.1063/1.2403866 (2006).
32  Castleton, C. W. M., Höglund, A. & Mirbt, S. Density functional theory calculations of defect energies using supercells. *Model. Simul. Mater. Sci. Eng.* **17**, 084003, doi:10.1088/0965-0393/17/8/084003 (2009).
33  Lin, S.-k. *et al.* Corrigendum to "Ab initio energetics of charge compensating point defects: A case study on MgO" [Comput. Mater. Sci. 73 (2013) 41–55]. *Computational Materials Science* **109**, 104, doi:10.1016/j.commatsci.2015.07.023 (2015).
34  Kaczmarowski, A., Yang, S., Szlufarska, I. & Morgan, D. Genetic algorithm optimization of defect clusters in crystalline materials. *Computational Materials Science* **98**, 234-244, doi:10.1016/j.commatsci.2014.10.062 (2015).
35  Large-scale Atomic/Molecular Massively Parallel Simulator (LAMMPS) (Sandia National Laboratories, 2004-2016).
36  Mayeshiba, T. T. & Morgan, D. D. Factors controlling oxygen migration barriers in perovskites. *Solid State Ionics* **296**, 71-77, doi:10.1016/j.ssi.2016.09.007 (2016).
37  Mayeshiba, T. & Morgan, D. Correction: Strain effects on oxygen migration in perovskites. *Physical chemistry chemical physics : PCCP* **18**, 7535-7536, doi:10.1039/c6cp90050j (2016).
38  Nam, H. O. & Morgan, D. Redox condition in molten salts and solute behavior: A first-principles molecular dynamics study. *Journal of Nuclear Materials* **465**, 224-235, doi:10.1016/j.jnucmat.2015.05.028 (2015).
39  Yu, M., Yankovich, A. B., Kaczmarowski, A., Morgan, D. & Voyles, P. M. Integrated Computational and Experimental Structure Refinement for Nanoparticles. *ACS Nano* **10**, 4031-4038, doi:10.1021/acsnano.5b05722 (2016).
40  *The MIT License*, <https://opensource.org/licenses/MIT> (2016).
41  Rong, Z., Kitchaev, D., Canepa, P., Huang, W. & Ceder, G. An efficient algorithm for finding the minimum energy for cation migraiton path in ionic materials. (2016).
42  yEd Graph Editor (https://www.yworks.com/products/yed, 2016).